\documentclass[aps,prl,reprint,superscriptaddress]{revtex4-1}
\usepackage{graphicx}
\usepackage[ansinew]{inputenc}
\usepackage{array}
\usepackage{color}
\usepackage{amsmath}
\usepackage{amsxtra}
\usepackage{amstext}
\usepackage{amssymb}
\usepackage{latexsym}

\begin{document}

\title{Optical characterization of electron-phonon interactions at the saddle point in graphene}

\author{Adam T. Roberts}
\affiliation{College of Optical Sciences, University of Arizona, Tucson, AZ 85721}
\affiliation{U.S. Army Aviation and Missile Research, Development, and Engineering Center, Redstone Arsenal, AL 35898}

\author{Rolf Binder}
\affiliation{College of Optical Sciences, University of Arizona, Tucson, AZ 85721}
\affiliation{Department of Physics, University of Arizona, Tucson, AZ 85721}

\author{Nai H. Kwong}
\affiliation{College of Optical Sciences, University of Arizona, Tucson, AZ 85721}

\author{Dheeraj Golla}
\affiliation{Department of Physics, University of Arizona, Tucson, AZ 85721}

\author{Daniel Cormode}
\affiliation{Department of Physics, University of Arizona, Tucson, AZ 85721}

\author{Brian J. LeRoy}
\affiliation{Department of Physics, University of Arizona, Tucson, AZ 85721}

\author{Henry O. Everitt}
\affiliation{U.S. Army Aviation and Missile Research, Development, and Engineering Center, Redstone Arsenal, AL 35898}

\author{Arvinder Sandhu}
\email[]{sandhu@physics.arizona.edu}
\affiliation{College of Optical Sciences, University of Arizona, Tucson, AZ 85721}
\affiliation{Department of Physics, University of Arizona, Tucson, AZ 85721}


\date{\today}

\begin{abstract}

The role of electron-phonon interactions is experimentally and theoretically investigated near the saddle point absorption peak of graphene. The differential optical transmission spectra of multiple, non-interacting layers of graphene reveals the dominant role played by electron-acoustic phonon coupling in bandstructure renormalization.  Using a Born approximation for electron-phonon coupling and experimental estimates of the dynamic phonon lattice temperature, we deduce the effective acoustic deformation potential to be $D^{\rm ac}_{\rm eff} \simeq 5$eV. This value is in accord with recent theoretical predictions but differs substantially from those obtained using electrical transport measurements.

\end{abstract}

\maketitle


Graphene research is fueling the development of new electronic and photonic devices such as high-speed field-effect transistors, efficient terahertz sources, ultrafast broadband photo-detectors, and photovoltaics.\cite{Novoselov2012,Bonaccorso2010}. Graphene offers ultrahigh charge carrier mobility, excellent heat conductivity, large photo-response bandwidth, among many other unique properties\cite{Novoselov2012}. Recent experimental and theoretical efforts have led to the recognition of the fact that carrier-carrier\cite{Neto2012} and carrier-phonon\cite{Kaasbjerg2012} couplings as well as excitonic effects  \cite{Heinz2011,Louie2009,malic-etal.11,groenqvist-etal.12}
can significantly alter the electronic bandstructure and optical properties of graphene. 

In particular, the electron-phonon interactions in graphene have numerous ramifications. The intrinsic carrier mobility in high quality graphene devices is limited by electron-phonon scattering\cite{Borysenko2010,Min2011,Bolotin2008,Chen2008,Dean2010}. An efficient opto-electronic device design also relies on the conversion of the energy of photoexcited carriers to electrical current before it dissipates through electron-phonon interactions\cite{Tse2009,Bistritzer2009,knorr2011b}. Ultrafast heat generation and dissipation dynamics in devices, which is an important topic in nanoscale heat management, is also crucially dependent on the interaction of electronic excitations with phonons\cite{Pop2010}.

Clearly, it is important to understand and quantify the many-body interaction effects in graphene, especially in the case of dynamically varying populations of phonons.
However, the exact nature and strength of electron-phonon (e-ph) coupling is unclear at present. Specifically, the electron-acoustic phonon interaction strength, characterized by the deformation potential $D^{\rm ac}_{\rm eff}$, has been controversial. The experimental estimates  obtained from electrical transport measurements\cite{Bolotin2008,Chen2008,Dean2010} range from $16-50$ eV, while theoretical predictions indicate acoustic deformation potential in the range of $\sim 2.8-7$ eV\cite{Kaasbjerg2012,Borysenko2010,low-etal.2012}. Since many observables are proportional to $|D^{\rm ac}_{\rm eff}|^2$, a $3-50$ eV range implies an uncertainty of more than two orders of magnitude.
Optical deformation potentials also show some variation; for example, Ref.\ \cite{low-etal.2012} uses 11 eV/\AA\ whereas Ref.\ \cite{sule-knezevic.12} uses 50 eV/\AA.

\begin{figure}[t]
\includegraphics[width=.5 \textwidth]{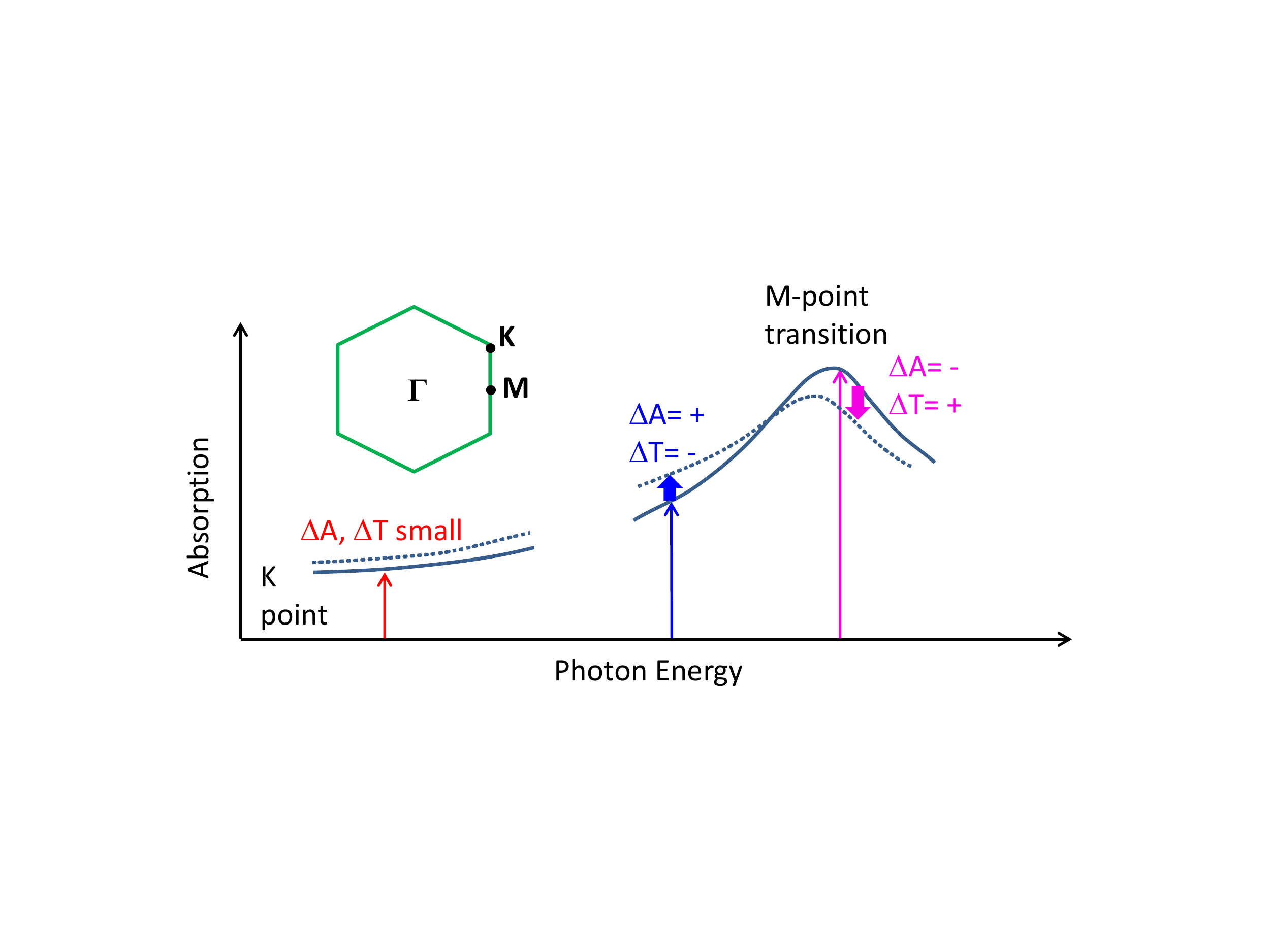}
\caption{ Schematic showing that the electron-phonon renormalization of bandstructure is easier to identify near the M point absorption maxima through a large change in the magnitude, accompanied by the change in sign of the energy-resolved differential transmission signal. Inset: Brillouin zone corresponding to the honeycomb lattice of graphene. The M-point corresponds to a saddle point in the electronic bandstructure.
}
\label{FigMpoint}
\end{figure}

Here we report ultrafast pump-probe spectroscopic measurements of the e-ph coupling in graphene as a sensitive alternative to electrical transport measurements \cite{Bistritzer2009}.  Instead of focusing on the heavily studied K-point in the graphene band structure \cite{Breusing2011,knorr2011a,Norris2011,Newson2009,Dawlaty2008,Sun2008}, our study uniquely concentrates near the M-point absorption peak which is much more sensitive to e-ph induced absorption band shifts and deformations than the relatively featureless absorption band near the K-point (Fig.\ \ref{FigMpoint}).  Our measurements constrain the acoustic deformation potential to values significantly lower than those obtained through electrical transport measurements but quite consistent with recent theoretical predictions.

The transmission modifications measured in our differential transmission spectroscopy detect only microscopic processes that have been modified by the pump pulse. This presents an advantage for the measurement of electron-acoustic phonon interactions, because the modification of e-ph scattering affects phonon absorption differently from phonon emission. For sufficiently small changes, phonon emission remains unchanged (as it is independent of the phonon occupation), whereas phonon absorption is proportional to the occupation numbers of phonons. At room temperature, optical phonons in graphene have much lower occupation numbers than acoustic phonons. If the pump-induced changes do not raise the temperature close to the optical phonon frequency $\hbar \Omega_{opt} (\simeq 200 {\rm meV})$, then the acoustic phonons play a dominant role in the observed signal. In our case, the estimated initial phonon temperature of $T=610$ K implies $k_BT \simeq 53 {\rm meV}$ which is substantially smaller than $\hbar \Omega_{opt}$. This regime allows us to probe the electron-acoustic phonon coupling and distinguishes our work from other measurement approaches such as angle-resolved photoemission spectroscopy (ARPES) \cite{Bostwick2007,Siegel2011}, which are more sensitive to optical phonon induced effects.

The graphene used in our study was grown through chemical vapor deposition\cite{Ruoff2009} and then transferred to a $200$ $\mu$m thick sapphire substrate. In order to enhance the signal level in our measurements, up to ten individually grown graphene layers were stacked on top of each other. The layers in the stack do not influence the monolayer behavior of the sample as was verified through Raman spectroscopy. Prior linear absorption measurements \cite{Yu2011} also indicate that there is minimal interlayer coupling in CVD graphene stacks compared to the exfoliated samples or multilayer films on SiC.

In our experiment, an ultrafast ($\sim$100fs) pump pulse photoexcites carriers, which relax through excitation of phonons\cite{knorr2011b}. The transmission of a time-delayed probe pulse is monitored over a wide range of photon energies on either side of the 4.6 eV M-point resonance, which is a local `saddle-point' absorption maximum\cite{Heinz2011}. After the pump-produced carriers have substantially relaxed through phonon emission, the interaction of probe-excited electrons with 
the phonon population manifests itself as a renormalization of the band-structure and modification of the probe absorption spectrum. Depending on the photon energy, the probe can thus experience a decrease or increase in the absorption (see Fig.\ \ref{FigMpoint}), and the change in the sign of differential transmission provides unambiguous evidence of band renormalization due to e-ph interactions.

\begin{figure}[t]
\includegraphics[width=.5 \textwidth]{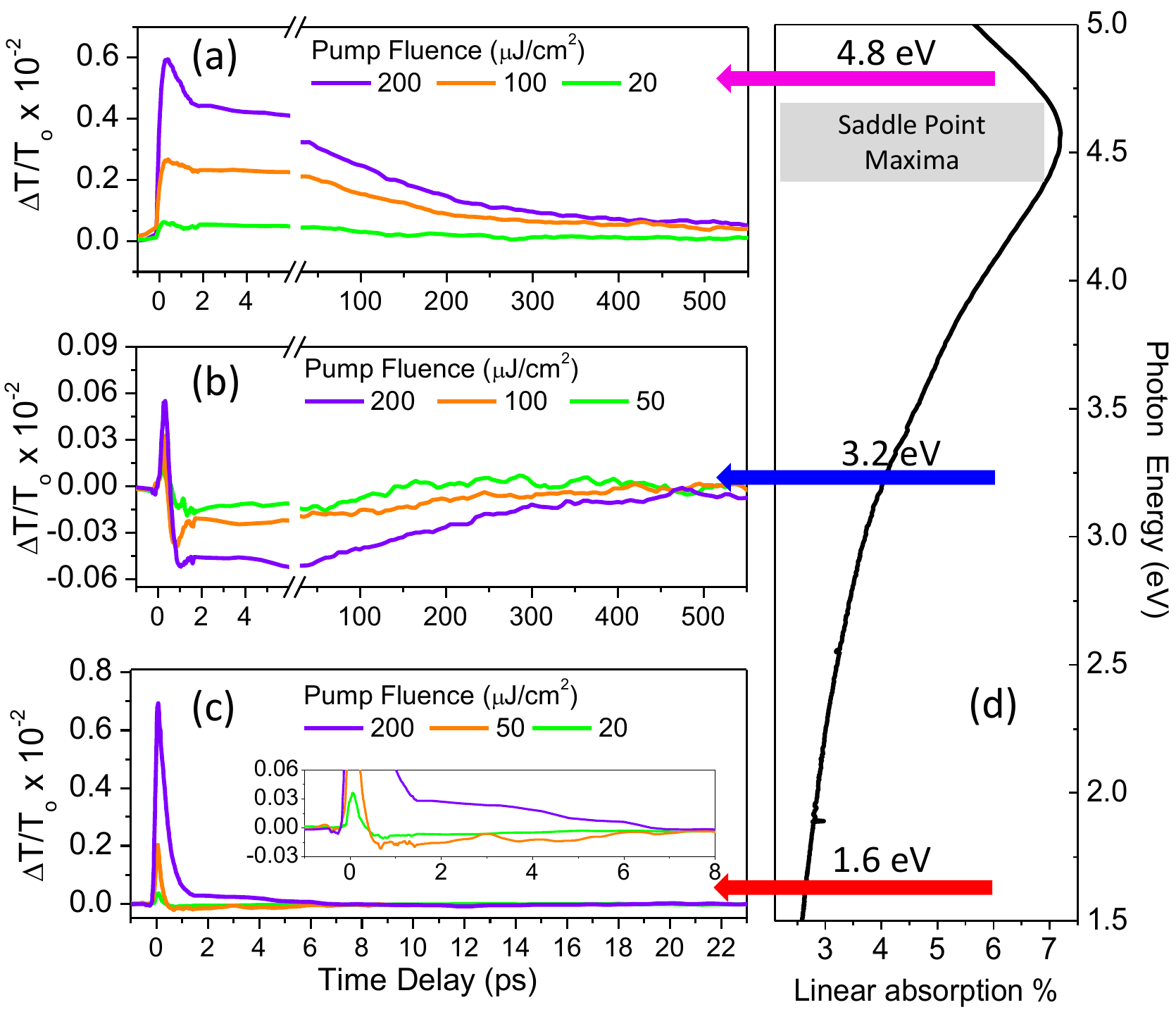}
\caption{ $\Delta T/T$ of the probe pulse at pump-probe energies of (a) 4.8 eV (b) 3.2 eV and (c) 1.6 eV. In each case different curves correspond to different pump fluence levels. The panel (d) shows the linear absorption profile indicating the probing energies relative to the absorption maxima at the saddle point.}
\label{time-resolved-DTT}
\end{figure}

The normalized pump-induced change in transmission ($\Delta T/T$) of a ten monolayer graphene stack is shown in Fig.\ \ref{time-resolved-DTT} for various photon energies on either side of the saddle point maxima. The measurements presented in this figure are degenerate in pump and probe energy (i.e. $\hbar\omega_{pu} = \hbar\omega_{pr}$). The pump fluence levels in Fig.\ \ref{time-resolved-DTT} correspond to carrier densities $\sim10^{12}-10^{13}$ cm$^{-2}$. Figures \ref{time-resolved-DTT}(a) and \ref{time-resolved-DTT}(b) correspond to photon energies of 4.8 and 3.2 eV respectively. Both figures show a positive transient lasting a few hundred femtoseconds, which can be attributed to Pauli blocking of the probe absorption due to the presence of pump-induced carriers in the conduction band\cite{Breusing2011}. The carriers quickly relax through carrier-carrier and carrier-phonon scattering mechanisms\cite{knorr2011b,Breusing2011,knorr2011a,Norris2011,Newson2009,Dawlaty2008,Sun2008,Shang2010}. After 2 ps, the plots in figures \ref{time-resolved-DTT}(a) and (b) exhibit a slow relaxation to the zero baseline. Notably, the sign of $\Delta T/T$ is opposite in these two cases. For the 4.8 eV case, we observe a strong positive $\Delta T/T$ signal, while, at 3.2 eV, it is negative. The negative $\Delta T/T$ indicates the generation of additional absorption channels which cannot be explained in the independent particle picture. The magnitude of the $\Delta T/T$ change is found to be proportional to the pump fluence (i.e. carrier density) in both cases. Importantly, the $\Delta T/T$ change lasts for the same duration in both cases ($\sim 300$ps), which suggests a common origin despite opposite signs. As discussed below, these observations can be attributed to the e-ph interaction induced band renormalization.

It is useful to compare 3.2 eV and 4.8 eV observations with the frequently studied case of $\hbar\omega_{pu} = \hbar\omega_{pr} =1.6$ eV. Our data at 1.6eV (Fig.\ \ref{time-resolved-DTT}(c)) shows a positive transient followed by a decay within 1ps, and a very small $\Delta T/T$ change beyond $t>2ps$. These observations are quite consistent with Pauli-blocking and carrier relaxation mechanisms discussed extensively in the literature\cite{Breusing2011,Norris2011,Newson2009,Dawlaty2008} and indicate that substantial phonon excitation and carrier recombination occurs within the first 2 ps. The transient signals beyond 2ps are thus expected to be dominated by phonon induced renormalization effect. However, as shown in the zoomed in view (inset in Fig.\ \ref{time-resolved-DTT}(c)), the positive or negative transients in the 1.6eV data are extremely weak beyond 2ps. Therefore near-infrared measurements ($\sim$1.6eV) in the Dirac cone region, where the linear absorption profile is essentially flat, do not permit clear identification of the e-ph interactions. In contrast, observations at higher photon energies exhibit stronger transients at long time delays, where e-ph interactions are dominant.

\begin{figure}[t]
\includegraphics[width=0.45 \textwidth]{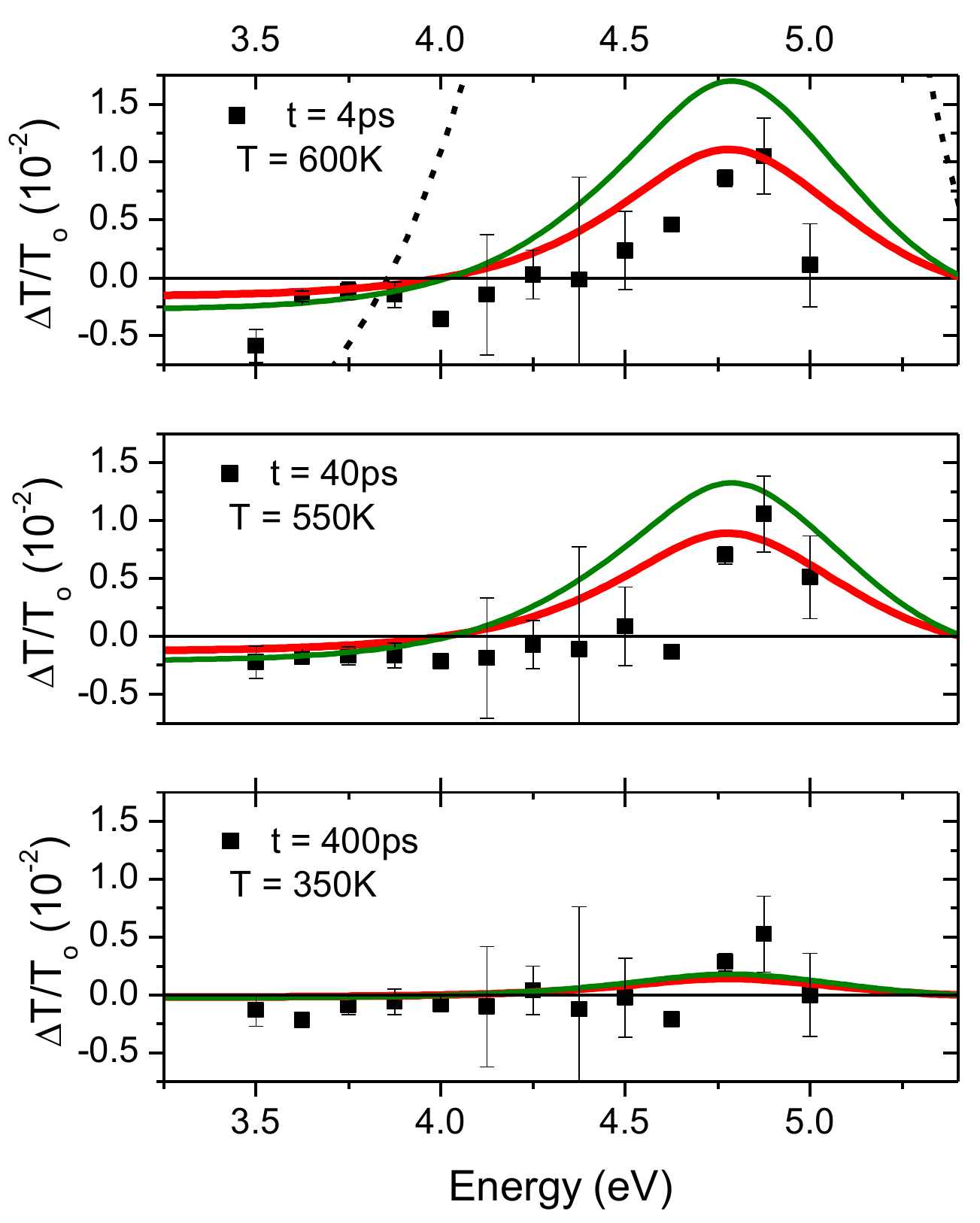}
\caption{ $\Delta T/T$ as a function of probe photon energy at three time delays (4, 40 and 400 ps). The pump pulse has fluence of 400 $\mu$J/cm$^{2}$ and fixed excitation energy of 4.8 eV. The calculated results for coupling to acoustic phonons only with $D^{\rm ac}_{\rm eff}=5.3$ eV are shown in red and acoustic+optical phonons with $D^{\rm op}_{\rm eff}=11$ eV/\AA\
are shown in green. The dashed black line shows acoustic-only coupling with $D^{\rm ac}_{\rm eff}=20$ eV. Each figure panel is labeled with specific phonon temperature values that are obtained using a procedure described in the text.}
\label{DTT-spectra}
\end{figure}

Specifically, we argue that the energy resolved measurements around the saddle point absorption maxima can serve to quantify electron-phonon coupling effects through large changes in the magnitude and accompanied by changes in the sign of $\Delta T/T$. In fact, hints of phonon induced band renormalization have been observed in some previous studies at high photon energies \cite{Shang2010,Seibert1990}. A detailed measurement of the `lineshape', i.e. the energy dependence of the $\Delta T/T$ near the saddle point energy at time delays greater than 2 ps, can thus provide a means to quantify the strength of e-ph interactions. We obtain the experimental lineshape of $\Delta T/T$, by fixing the pump excitation energy to 4.8 eV and varying the probe photon energy around the saddle point transition. These measurements are shown in Fig.\ \ref{DTT-spectra} for time delays of 4, 40, and 400 ps.

To understand the microscopic origin of many-body effects observed in our $\Delta T/T$ measurements we utilize a real-time nonequilibrium Green's function technique that allows us to compute band renormalizations due to many-body interactions. We model the expected lineshape of $\Delta T/T$ near the saddle point transition for e-ph interactions. Using the Born approximation for the e-ph interaction, the single-phonon selfenergy $\Sigma_s ({\bf k})$, where $s=\pm 1$ denotes the  $\pi, \pi^{\ast}$ bands and ${\bf k}$ the wavevector, assumes the standard form (given, e.g., by Eq.\ (3.5.10) in \cite{mahan.86}) except that we keep the band dependence of the electron energies $E_s({\bf k})$ (see Eq.\ 6  in \cite{castroneto-etal.09} with $t'=0$), sum over  electron band and phonon indices, and restrict the sum to the first Brillouin zone. Our squared matrix element for acoustic phonon coupling reads
$
\left|  M_{s,s'}^{\rm ac}  ({\bf k, k'}) \right|^2 = \frac{\hbar {\cal A}_o }{4 m  \omega_{\bf q}^{\rm LA}} \left| D^{\rm ac}_{\rm eff} q a \langle u_{s {\bf k}} | u_{s' {\bf k}'} \rangle
\right|^2
$
where the longitudinal acoustic phonon  frequencies are denoted by $\omega_{\bf q}^{\rm LA} $ (all phonon frequencies are obtained numerically after Ref.\ \cite{suzuura-ando.02}), $m$ is the carbon nucleus mass,
$a$ the bond length,
${\cal{A}}_0$ the area of the unit cell, ${\bf q} = {\bf k - k'}$ reduced to first Brillouin zone, and
$| u_{{\bf k}' , s} \rangle$ are the lattice-periodic parts of the two-component spinor Bloch function.
An effective deformation potential with coupling to longitudinal phonons (denoted by ${\Xi}_{\rm eff}$ in Ref.\ \cite{Kaasbjerg2012}) has been found to provide a quantitative model for the combined coupling to longitudinal and transverse acoustic phonons. Following the discussion on page 6 of Ref.\ \cite{Kaasbjerg2012}, we set $D^{\rm ac}_{\rm eff} $ = $5.3$ eV, and although our measurement is at the M point, we expect $D^{\rm ac}_{\rm eff}({\rm K}) \simeq D^{\rm ac}_{\rm eff}({\rm M})$ \cite{footnote-D-kdependence}.
For coupling to optical phonons, we proceed in the same semi-phenomenological manner to introduce an effective optical deformation potential $D^{\rm op}_{\rm eff}$ with the coupling matrix element
$
\left|  M_{s,s'}^{\rm op}  ({\bf k, k'}) \right|^2 = \frac{\hbar {\cal A}_o }{4 m \hbar \omega_{\bf q}^{\rm LO}} \left| D^{\rm op}_{\rm eff}  a
\langle u_{s {\bf k}} | u_{s' {\bf k}'} \rangle
\right|^2
$.
Our model considers only the coupling of electrons to in-plane phonon modes because out-of-plane phonon modes are expected to have a negligible contribution\cite{Mariani2008,Borysenko2010}.
The selfenergy $\Sigma_s ({\bf k})$ depends on the pump-probe delay time via the time dependent
phonon occupation functions.

The spectra are calculated from the susceptibility
of a graphene layer which is defined as
$
\chi(\omega) =
\sum_{\bf k} R({\bf k})
{\cal{L}}[\omega,\Delta E,\gamma]
L_F({\bf k})
$
, where
$R({\bf k}) = (-2 e^2 / (m^2 \omega^2 A)) \langle 
\left[
{\bf p} ({\bf k}) \cdot \hat{{\bf e}}_0
 \right]^2
\rangle_{\theta}$,
$A$ is the graphene area, $e$ ($m$) the electron charge (mass), {\bf p}({\bf k})  the  momentum matrix element, $\hat{{\bf e}}_0 = ( \cos \theta,
\sin \theta)$  the polarization unit vector of the light field (we average
over  $\theta$),
${\cal{L}}[\omega,\Delta E ,\gamma ] = (\omega - \Delta E +i \gamma)^{-1}$,
$\Delta E  = \Delta E_0 ({\bf k})
+ {\rm Re} ( \Delta \Sigma_1 ({\bf k}) -  \Delta \Sigma_{-1} ({\bf k}))$ the renormalized transition energy from the $s=-1$ to the $s=1$ band,
$L_F({\bf k})$ a phenomenological Fano-like lineshape factor after Ref.\ \cite{Heinz2011},  and
$\gamma  = \gamma_0(\omega) - {\rm Im } ( \Delta \Sigma_1({\bf k}) -  \Delta \Sigma_{-1}({\bf k}))$, 
where 
$\gamma_0(\omega) = \gamma_0 {\rm tanh}(5  \omega / \omega_b )$
($\gamma_0 = 0.8$ eV,  $\hbar \omega_b = 4.6$ eV) is a phenomenological broadening.
In the unrenormalized transition energies $\Delta E_0 ({\bf k})$, we use a hopping parameter ${t} = 2.38$ eV in order to incorporate the excitonic shift of the absorbance maximum such that it matches the experimental peak. We use ${t} = 2.55$ eV \cite{samsonidze2007} in the calculation of self-energies.
The susceptibility is used in a conventional transfer matrix approach that determines the transmitted and reflected light field amplitudes.

To compare the theoretical calculation of this lineshape with the observed experimental data, the phonon temperature $T(t)$ must be estimated at various time delays. We estimate the initial temperature $T_0=T(t_0)$ by assuming that within a short time $t<2$ ps\cite{Breusing2011,knorr2011b, Norris2011,Newson2009,Dawlaty2008,Shang2010}, the photoexcited carriers relax and the supplied laser pulse energy $E_L$ is transferred to the phonons with internal energy given by
$U_{ph}(T_0) = \sum \limits_{{\bf q} \mu} \hbar \omega^{\mu}_{{\bf q}} [n_\mu^{\rm ph} ( {\bf q}, T(t_0)) +1/2 ]$. Here $\mu$ runs over all in-plane phonon branches, and $n_\mu^{\rm ph}$ denotes the Bose phonon occupation function.
By calculating the internal energy of phonons as a function of temperature, the initial temperature $T_0$ is  found using $U_{ph}(T_0) = U_{ph}(300 {\rm K}) + E_L$. For the highest incident fluence of $400$ $\mu$J/cm$^2$, the estimated initial temperature of the phonon system is 610 K. The temperature will decrease as the phonons decay and transfer energy to the substrate. We find that in the e-ph coupling dominated region ($t>2$ ps), the $\Delta T/T$ at 4.8 eV is linearly related to the temperature, so the temperature decay profile may be estimated by fitting an exponential decay curve to $\Delta T/T$ at 4.8eV in figure \ref{time-resolved-DTT}(a).

Using this method, the phonon temperatures are estimated to be 600 K, 550 K, 350 K at time delays of 4, 40 and 400 ps respectively.
The numerical calculations using the predicted \cite{Kaasbjerg2012} $D^{\rm ac}_{\rm eff}=5.3$ eV reproduce the measured lineshape, the $\Delta T/T$ maxima, and the zero-point crossing quite well for all three time delays (Fig.\ \ref{DTT-spectra}, red curves).
Note that $D^{\rm ac}_{\rm eff}=20$ eV, a value typically obtained in electrical transport measurements, yields an inordinately large $\Delta T/T$ (dashed line, top panel of Fig.\ \ref{DTT-spectra}) inconsistent with our optical measurements.  Good agreement between theory and experiment is obtained by including only electron-acoustic phonon coupling. Adding the optical phonon contribution increases the signal slightly, but leaves the lineshape unchanged. Using the $D^{\rm op}_{\rm eff}$ = 11 eV/\AA\ from Ref.\ \cite{low-etal.2012} still yields reasonable agreement with experiment, especially at 40 and 400 ps (Fig.\ \ref{DTT-spectra}, green curves).

 Despite overall good agreement between experiment and theory, the calculated zero-crossing energy in $\Delta T/T$ lineshape of 4.1 eV  differs slightly from the experimentally measured value of 4.3 eV (Fig.\ \ref{DTT-spectra}). Since the zero-crossing point is an important and stable feature of the experimental lineshape, it could be used as a guide for further improvements in the theoretical model or prediction of new effects. For example, if we imagine a possibility where a small population of pump-induced carriers survives for a long time delay, then electronic interactions can also play a role. We have conducted preliminary calculations of electron-electron interactions to consider such a possibility. Using the statically screened Hartree-Fock approximation from Eq.\ (9.29) of \cite{haug-koch.04}, extended to account for both inter- and intra-band Coulomb matrix elements in graphene, we find that a combination of the two effects - a dominating phonon contribution and weak electronic contribution - can blue shift the calculated zero crossing point. However, fairly high electron densities ($\sim$ 10$^{12}$ cm$^{-2}$) would be required to produce a significant shift. Interestingly, the $\Delta T/T$ lineshape resulting solely from electron-electron interactions is completely different from the lineshape resulting from electron-phonon interactions. The lineshape measurement could thus be generally used for identifying the nature of many-body interactions in various situations. Extensions and improvements to the theory, such as microscopic modeling of exciton-phonon coupling, might further reduce the small discrepancies between theoretical and experimental lineshapes.

In summary, we present differential transmission spectra of graphene over a wide range of frequencies in the vicinity of the M saddle point. 
The observed lineshape of the differential transmission is consistent with a many-particle theory using an effective electron-acoustic-phonon deformation potential of  $D^{\rm ac}_{\rm eff} \simeq 5$ eV. Future time-resolved studies using nonlinear spectroscopy at the M-point could be aimed at investigation of additional physical processes such as fast carrier relaxation dynamics occuring on femtosecond timescales and the role of excitonic-phonon coupling.

We thank John V. Foreman for assistance with experiments. A.R. gratefully acknowledges support from the DoD SMART program. D.C. and B.J.L. acknowledge support from the U.S. Army Research Laboratory under grant number W911NF-09-1-0333.

\bibliography{RefsGraphene}

\end{document}